\documentclass{cs19proc}

\graphicspath{{Images/}}

\editors{G.~A. Feiden}
\publisher{Zenodo}
\conference{The 19th Cambridge Workshop on Cool Stars, Stellar Systems, and the Sun}
\conferencedate{2016}

\title{2D dynamics of the radiation zone of low mass stars}
\author{Delphine Hypolite$^{1}$,
        St\'ephane Mathis$^{1}$,
        Michel Rieutord~$^{2}$
        }

\affiliation{$^{1}$ Laboratoire AIM Paris-Saclay, CEA/DRF - CNRS - Universit\'e Paris Diderot, IRFU/SAp Centre de Saclay, F-91191 Gif-sur-Yvette Cedex, France\\
			 $^{2}$ Institut de Recherche en Astrophysique et Plan\'etologie, Observatoire Midi-Pyr\'en\'ees, Universit\'e de Toulouse, 14 avenue Edouard Belin, 31400 Toulouse, France}

\shorttitle{2D dynamics of the radiation zone of low mass stars}
\shortauthors{D. Hypolite, S. Mathis \& M. Rieutord}

\abs{The internal rotation of low mass stars all along their evolution is of primary interest when studying their rotational dynamics, internal mixing and magnetic fields generation. In this context, helio- and asteroseismology probe angular velocity gradients deep within solar type stars. Still the rotation of the close center of such stars on the main sequence is hardly detectable and the dynamical interactions of the radiative core with the surface convective envelope is not well understood. Among them, the influence of the differential rotation profile sustained by convection and applied as a boundary condition to the radiation zone may be very important leading to the formation of tachoclines. In the solar convective region, the equator is rotating faster than the pole while numerical simulations predict  either a solar or an anti-solar rotation in other low mass stars envelopes depending on their convective Rossby number. 
In this work, we therefore build for the first time 2D steady hydrodynamical models of low mass stars radiation zone providing a full 2D description of their dynamics and studying the influence of a general shear boundary condition accounting for a solar or an anti-solar differential rotation in the convective envelope. We compute coherently differential rotation and the associated meridional circulation using the Boussinesq approximation.}

\begin{document}

\maketitle

\section{Introduction}

It has long been known that rotation plays a  central role in the dynamical and chemical evolution of stars (\cite{meynetmaeder00}, \cite{maedermeynet09}).
Through the rotationnal mixing driven by the differential rotation, the meridional circulation  and the turbulence it sustains, chemical elements and angular momentum are transported through the radiation zones of stars (e.g. \cite{zahn92}, \cite{maederzahn98}, \cite{mathiszahn04}, \cite{MR06}, \cite{ELR07}) in a way that needs to be precisely characterized  and modeled.

\begin{figure}
	\centering
	\includegraphics[trim = 2cm 3.5cm 2cm 3.5cm,width=0.8\linewidth,angle=-90]{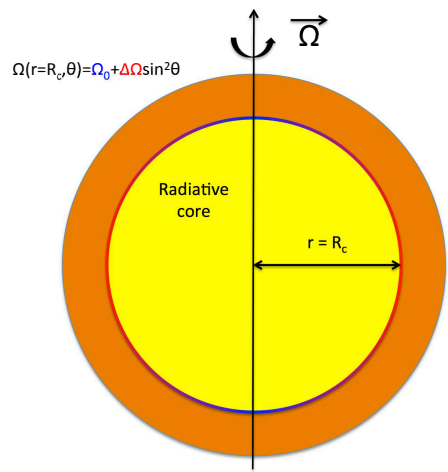}
	\caption{Sketch of a radiative core (yellow region) upon which lies a differentially rotating convective envelope (orange region); $R_c$ is the radius of the radiative core. At the interface $r=R_c$, the rotation rate is imposed: $\Omega_{cz}(r=R_c,\theta)= \Omega_0 + \Delta\Omega\sin^2\theta$ where $\Omega_0$ is the rotation rate of the pole of the radiative core, i.e. $\Omega_0=\Omega(r=R_c,\theta=0)$. The equator rotates faster than the pole when $\Delta\Omega>0$ and slower than the pole when $\Delta\Omega<0$. \label{fig1} 
	}
\end{figure}
\begin{figure}
	\centering
	\includegraphics[width=1\linewidth]{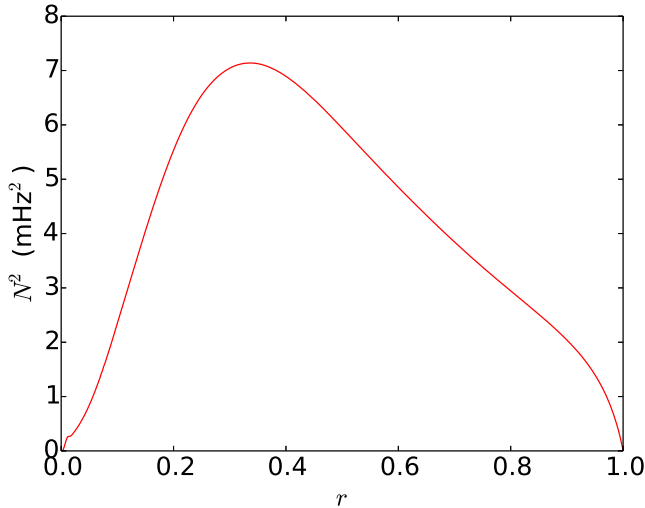}
	\caption{Squared Brunt-V\"ais\"al\"a frequency from a 1D $1 M_\odot$ MESA model near the ZAMS (with $Z=0.02$, $\alpha_{\rm MLT}=2$).}	\label{fig2}
\end{figure}
Indeed, helioseismology provides the internal rotation profile deep within the Sun (until $0.2R_\odot$) where the radiation zone has a quasi solid rotation under the tachocline (\cite{couvidat03}, \cite{garcia07}) which 1D rotating stellar models fail to reproduce (\cite{TC05}, \cite{TC10}).
One reason probably lies in the need for other physical mechanisms such as internal gravity waves (\cite{zahn97}, \cite{TC05}) or an internal magnetic field (\cite{GM98}, \cite{spruit99}) or/and in the multidimensional nature of the differential rotation and of the meridional circulation \cite[][]{MR06}. 
Moreover, one wants to go beyond the slow rotation hypothesis underlying the 1D modeling of rotating stellar radiation zones to properly describe the dynamics occuring during the early stages of the evolution of stars during which they are rapidly rotating (\cite{GB13}, \cite{GB15}).
To tackle this issue, we build a 2D numerical model of a fast rotating sphere representing such a radiative core at the top of which we impose a latitudinal shear so as to reproduce the conical differential rotation applied by the convective zone. 

Numerical simulations (\cite{matt11}, \cite{kapyla14}, \cite{gastine14}, \cite{varela16}) reveal that the differential rotation of the convective envelope of low mass stars may be solar (with slow pole and fast equatorial regions) or anti-solar (with fast pole and slow equator) depending on the value of the convective Rossby number, which quantifies the ratio of the rotation period and of the convective turnover time (a small Rossby number corresponds to the rapidly rotating regime; see e.g. \cite{brun15}). We propose to perform a systematic parameter study over the boundary condition describing the convective differential rotation applied at the top of the radiation core between the pole and the equator. 
We provide a full 2D description of the differential rotation and the associated meridional circulation computed self-consistently using the Boussinesq approximation.
Since previous 2D modelings  (\cite{friedlander76}, \cite{garaud02}) are limited to the solar case, this setting lays a first important step for the next generation of rotating models of low mass stars in 2D.

\section{Description of the model}

\subsection{Scaled equations of the dynamics}

We consider a viscous fluid enclosed within a spherical shell, see Fig. \ref{fig1}, and aim to describe the velocity field and the temperature field resulting from the combined action stable stratification and rotation. 
The core is rotating and we assume that it is not deformed, meaning that the effects of the centrifugal acceleration are neglected. 
We focus on the case of fast rotation, which leads to the damping of the baroclinic modes on a  time scale shorter than in the slow rotation case \cite[][]{busse81}.
It allows us to describe the steady state which arises from this configuration and which is the solution  
of the system of equations (6) from \cite{MR06} (hereafter R06) in the Boussinesq approximation, which reads 
\begin{equation}
\left \{
\begin{array}{lcl}
\vec{\nabla} \times (\vec{e}_z \wedge \vec{u} - \theta_T \vec{r} - E\Delta\vec{u}) = - n^2(r) \sin{\theta}\cos{\theta}\vec{e}_{\varphi}\; ,\\
\left(\frac{n^2(r)}{r} \right)u_r = \frac{E}{Pr}\Delta\theta_T\; ,\\
\vec{\nabla}\cdot\vec{u} = 0\; ,\\
\end{array}
\right.
\label{eq1}
\end{equation}
where $\vec{e}_z$ is the axis of rotation of the fluid, 
$\vec{u}$ is the scaled velocity field using the baroclinic scale $V=\frac{\Omega_0\mathcal{N}^2R_c^2}{2g_S}$ from R06 where $\mathcal{N}^2=\alpha T_\star g_S/R_c$ is the scale of the squared Brunt-V\"ais\"al\"a frequency profile, $\alpha$ is the fluid dilation coefficient and $g_S$ is the surface gravity. 
The dimensionless temperature field $\theta_T$ is scaled using $\frac{\Omega_0^2R_c}{g_S}T_\star$ as a temperature scale with $T_\star$ an arbitrary stellar temperature scale.
In the same way, the scaled squared Brunt-V\"ais\"al\"a frequency profile $n^2(r)$ using is scaled $\mathcal{N}^2$. 
We choose to work within the corotating frame with the pole of the shell rotating at the rate $\Omega_0$.
The first equation is known as the vorticity equation. 
Its first term is the Coriolis acceleration, the second one is the buoyancy force, the third one the viscous force and on the right hand side is the baroclinic torque arising in stably stratified fluid.
The second equation is the energy equation where the heat advection balances its diffusion and where we do not take into account as a first step any nuclear heating. 
The last equation is the continuity equation when using the Boussinesq approximation.

The dimensionless numbers that characterize such a system are:
\begin{itemize}
\item The Ekman number $E=\displaystyle{\frac{\nu}{2\Omega_0R_c^2}}$.

It quantifies the importance of the kinematic viscosity $\nu$ over the Coriolis effects where $R_c$ is the radius of the radiative core. Using realistic values from ZAMS 1D MESA stellar models (\cite{paxton10}) for the microscopic viscosity and the radius $R_c$, we show that in the solar case ($M=1M_\odot$, $Z=0.02$, $\alpha_{\rm{MLT}}=2$), the Ekman number is around $\sim10^{-11}-10^{-13}$.
This value is very small and numerical simulations cannot reach it. We therefore study the case $E=10^{-6}$, which still describes qualitatively the physical behavior of the solution (the asymptotic regime is reached).

\item The Prandtl number $Pr=\displaystyle{\frac{\nu}{\kappa}}$.

This number is the ratio of the viscosity to the thermal diffusivity.
We use the solar value $Pr=2.10^{-6}$ (see e.g. \cite{ruediger14}) for our simulations.

\end{itemize} 

\begin{figure*}[ht]
	\centering
	\includegraphics[width=0.3\linewidth]{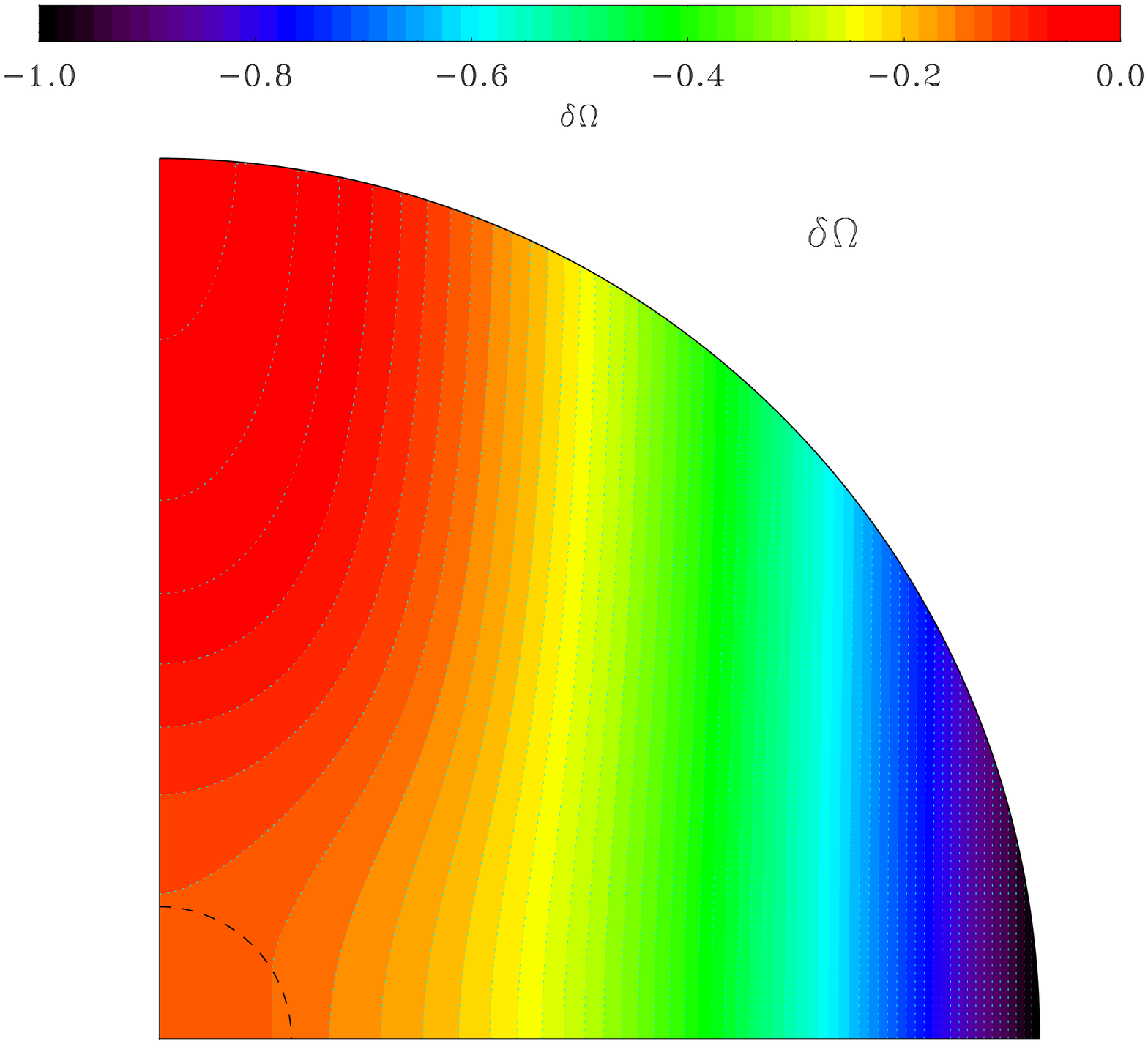}
	\includegraphics[width=0.3\linewidth]{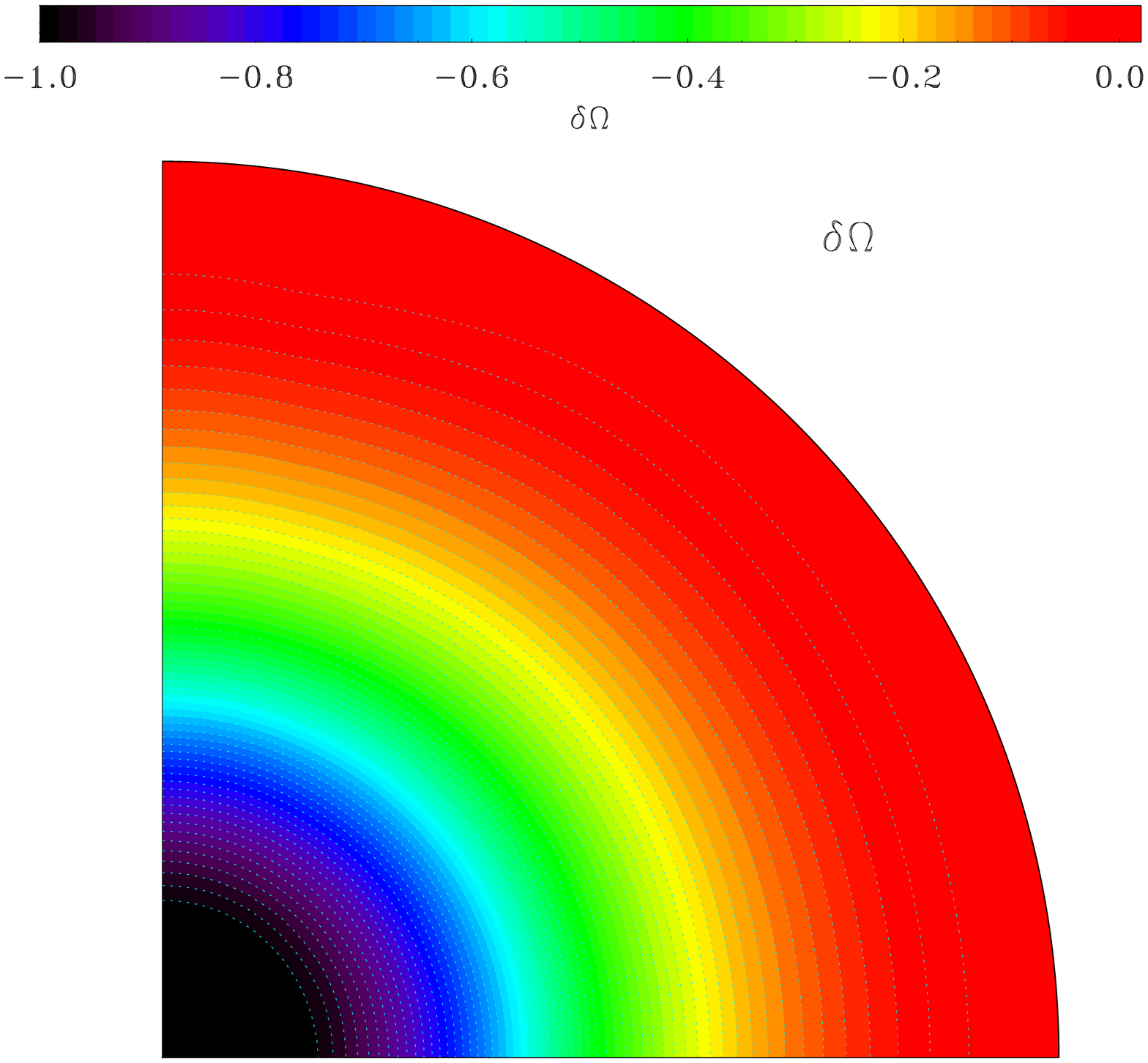}
	\includegraphics[width=0.3\linewidth]{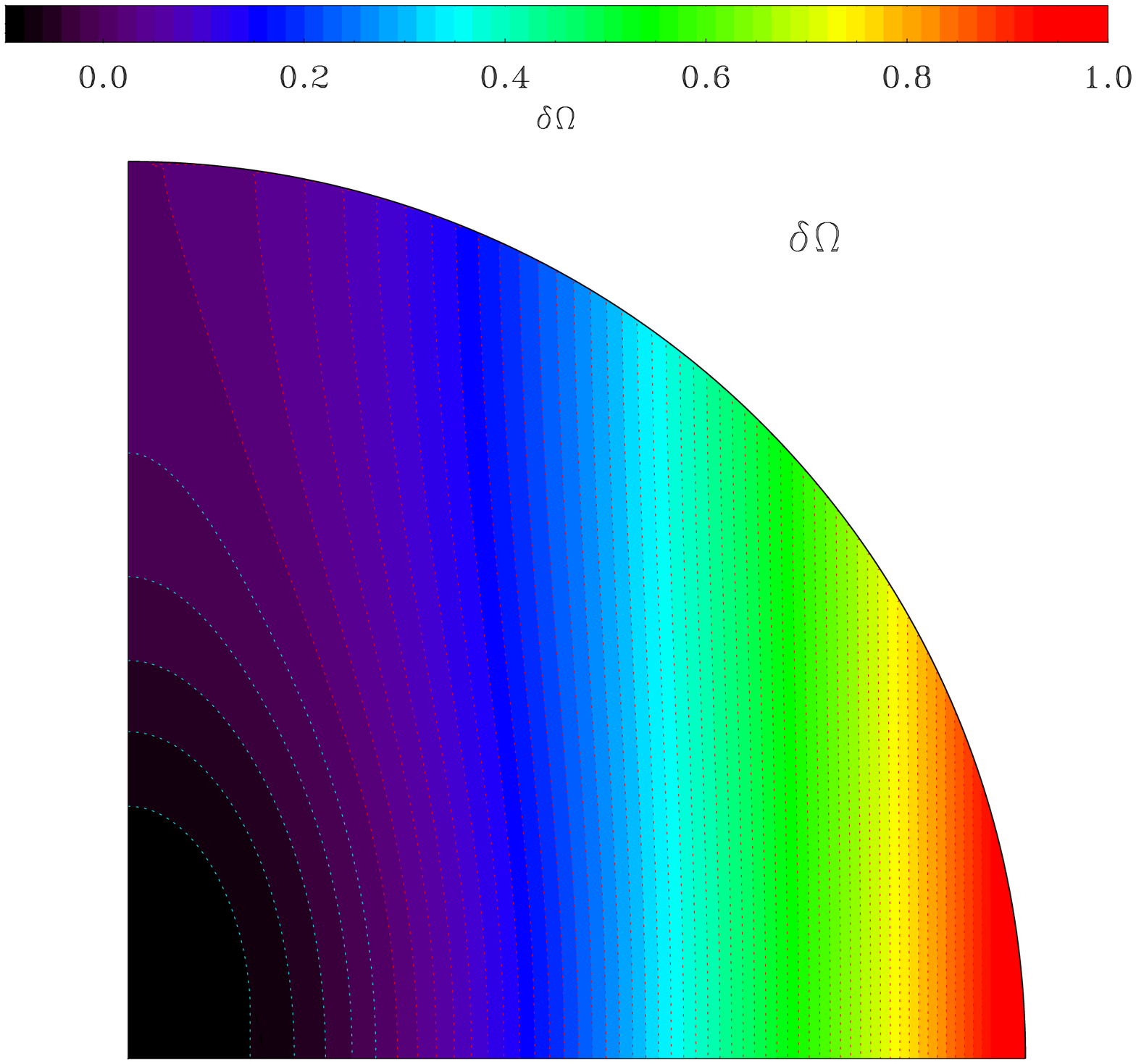}	
	\caption{Differential rotation $\delta \Omega$ shown in the meridional plane for $E=10^{-6}$, $\Pr=2.10^{-6}$ when using the Boussinesq approximation for $b=\{-10,10^{-2},10\}$ (from left to the right).  The stellar rotation axis is along the vertical.}
	\label{fig3}
\end{figure*}
\begin{figure*}[ht]
	\centering
	\includegraphics[width=0.3\linewidth]{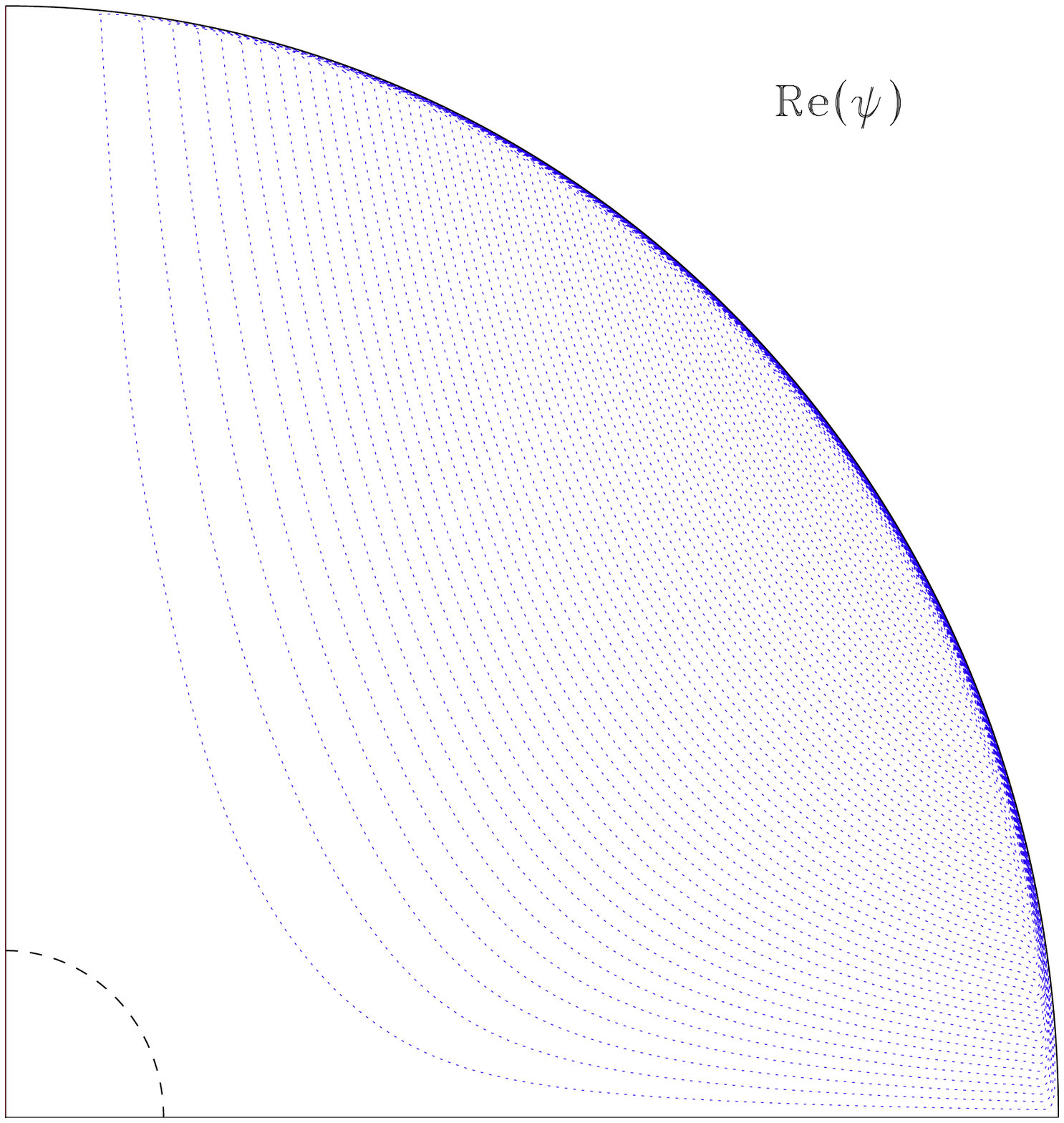}
	\includegraphics[width=0.3\linewidth]{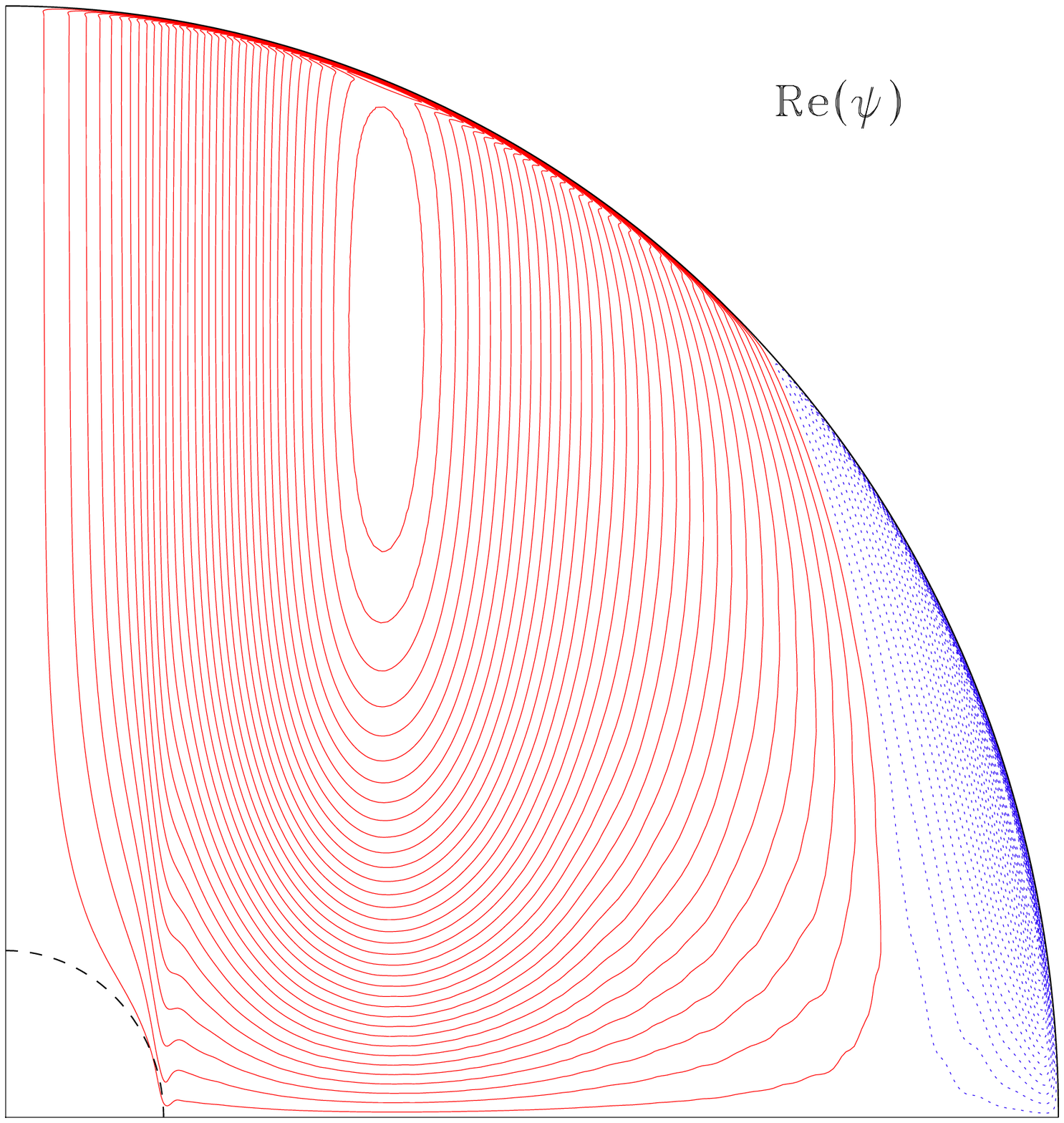}
	\includegraphics[width=0.3\linewidth]{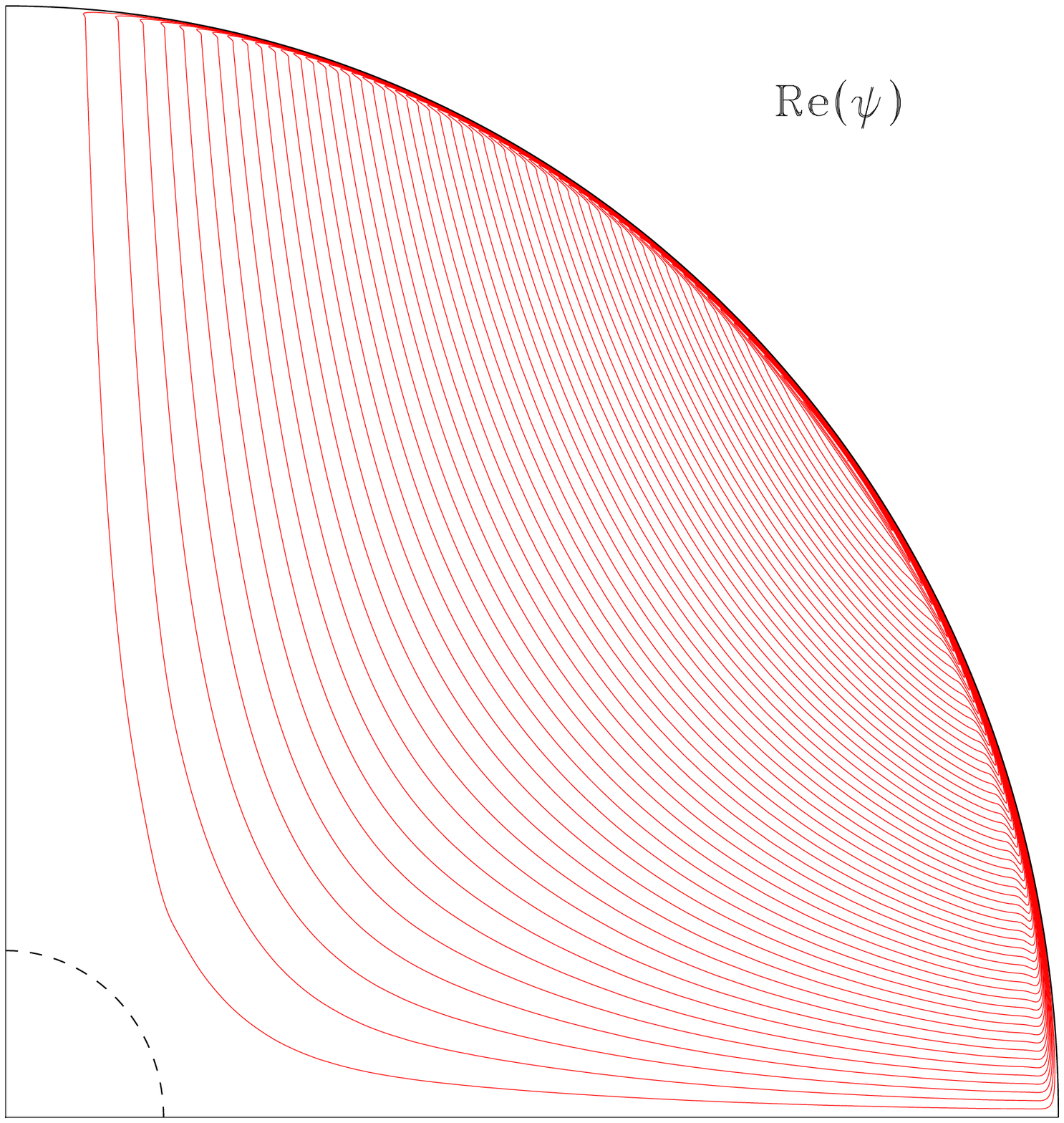}	
	\caption{Meridional circulation stream function $\psi$ (red : direct sens, blue: clockwise sens)  shown in the meridional plane for $E=10^{-6}$, $\Pr=2.10^{-6}$ when using the Boussinesq approximation for $b=\{-10,10^{-2},10\}$ (from left to the right).  The stellar rotation axis is along the vertical.}
	\label{fig4}
\end{figure*}

\subsection{Dynamical boundary conditions}

We impose a shear at the surface $r=R_c$ in order to account for a differentially rotating convective zone lying at the top of the radiation core such as 
\begin{equation}
\Omega_{cz}(r=R_c,\theta)= \Omega_0 + \Delta\Omega\sin^2\theta\; ,
\end{equation}
which is the simplest expression that simulations of the dynamics of stellar convective envelopes inspire (\cite{matt11}, \cite{kapyla14}, \cite{gastine14}).
When using dimensionless quantities in the corotating frame with the pole of the radiation core, the azimuthal velocity reads
\begin{equation}
u_\varphi(r=1,\theta)= b \sin^3\theta\; .
\end{equation}
The dimensionless number $b=\frac{R_c \Delta\Omega}{V}=\frac{1}{\mathcal{R}o}\frac{\Delta\Omega}{2\Omega_0}$ quantifies the shear at the top of the radiation core. Differential rotation is solar-like when the equatorial regions rotate faster than the pole ($b>0$) and anti-solar otherwise ($b<0$).
The Rossby number has been written $\mathcal{R}o=\frac{V}{2\Omega_0R_c}$ and quantifies the amplitude of the non-linearity of the system. This number is always less than $10^{-2}$ in the solar radiation core and therefore allows to focus on the linear case as long as $b<10^2$.

The meridional components of the velocity fields are set to zero at the upper boundary $r=R_c$ since the convection envelope applies stresses at the top of the radiative core and act as a non-penetrative condition.
At the center of the core ($r=0$), 
the azimuthal velocity, 
the radial velocity, its first radial derivative
and the first radial derivative of the temperature perturbation are set to zero.

In the next section, we perform a systematic parameter study over the shear parameter $b$.

\section{Numerical results}

We use a spectral method to solve numerically the set of Eqs. (\ref{eq1}). 
On the horizontal directions, the solution is described by vectorial spherical harmonics on the $(\vec{R}_l^m,\vec{S}_l^m,\vec{T}_l^m)$ basis \cite[][]{R87}
where
\begin{equation}
\vec{R}_l^m=Y_l^m(\theta,\varphi) \vec{e_r},\qquad \vec{S}_l^m=\vec{\nabla}_H Y_l^m, \qquad \vec{T}_l^m=\vec{\nabla}_H \wedge \vec{R}_l^m\; .
\end{equation}
$Y_l^m$ are the normalized spherical harmonics, functions of colatitude and azimuth ($\theta$,$\varphi$), $\vec{e_r}$ is the unit radial vector and the horizontal gradient $\vec{\nabla}_H=\partial_\theta \vec{e_\theta}+\frac{1}{\sin\theta}\partial_\varphi \vec{e_\varphi}$ is defined on the unity sphere.

Radially, the solution is projected on Chebyshev polynomials on a Gauss-Lobatto grid (e.g. \cite{canuto06}). This grid has more points on the edge allowing a good description of the boundary layer at the top of the domain.

We here use a profile from a ZAMS\footnote{We use the stopping condition \textit{stop\_near\_zams} of the MESA code to produce this model (http://mesa.sourceforge.net).}  $1M_\odot$ 1D MESA model (with $Z=0.02$, $\alpha_{\rm MLT}=2$) for interpolating the Brunt-V\"ais\"al\"a frequency profile on the Gauss-Lobatto grid. 
We show this profile within the radiation zone of the model on Fig. \ref{fig2}.
Changing the mass considered within the range of low mass stars ($0.5-1.1M_\odot$) does not affect the behavior of the solutions presented next.

On Fig. \ref{fig3}, we compute and show the resulting differential rotation $\delta\Omega$ relative to the pole rotation for $b=\{-10, 10^{-2}, 10\}$. The extremum value normalizes the field in each case.
When $|b|\leq 10^{-2}$, the solution is the baroclinic solution described by R06. The shear is too weak to be felt. This solution is called the thermal wind with a roughly shellular differential rotation close to the equator.
When $|b|>10^{-2}$, the amplitude of the geostrophic flow arising from the shear applied at the upper boundary is larger than the baroclinic solution and the Taylor-Proudman balance tends to be restored (quasi columnar structure).
Indeed, the azimuthal velocity at the equator has an $\mathcal{O}(b)$ amplitude when we subtract the baroclinic solution and we show numerically that the baroclinic solution is overcomed for $|b|>10^{-2}$. 
The associated differential rotation thus tends towards a cylindrical profile.

On Fig. \ref{fig4}, the meridional circulation is computed for $b=\{-10, 10^{-2}, 10$\}. 
When the baroclinic solution dominates, the number of cells of the meridional circulation is equal to the number of inflection points of the Brunt-V\"ais\"al\"a frequency profile plus one, here two. 
The rotation aligns the cells with the cylindrical z-direction (i.e. along the axis of rotation).
When $|b|>10^{-2}$, the dynamics is dominated by the geostrophic flow arising from the shear and the meridional circulation is dominated by a single, global circulation cell in each hemisphere.
For $b$ positive (solar-like differential rotation), the meridional circulation is counterclockwise and clockwise for negative $b$.
At the radiation/convection interface, the fluid moves toward the pole which rotates slower than the equator (for $b>0$ and vice-versa for $b<0$).

\section{Discussion and perspectives}

In this work, we propose a first 2D description of the dynamics of fast rotating radiative cores of low mass stars undergoing the shear induced by the differential rotation of the convective envelope at its upper boundary.

This simple Boussinesq model shows the necessity to resort to a 2D approach. Indeed, the arising flow from the shear is cylindrical, which must be described using a large number of spherical harmonics.
Such a setting also permits to take into account the Brunt-V\"ais\"al\"a frequency from selected evolutionary stages and stellar masses allowing the exploration of the HR diagram in 2D.

In the solar case, the shear is evaluated close to $b_\odot\simeq10$.
Therefore, the third plot of Fig. \ref{fig3} exhibits the best fast rotating solar model we may provide within the current setting.
We compute a core to surface rotation rate ratio lower than $1$ which is in agreement with what \cite{benomar15} deduced from observations.
But it also tells us to expect the differential rotation to be cylindrical until half of the radiative core depth with a fast equatorial region.
If we compare to the actual solar rotation profile, this calls for other physical processes responsible for transport of angular momentum deep within the internal regions, which is not taken into account in the present setting. 
Internal gravity waves \cite[][]{zahn97}, anisotropic turbulence (\cite{zahn92}, \cite{maeder03}, \cite{mathiszahn04}) and magnetic fields (\cite{GM98}, \cite{spruit99}, \cite{strugarek11}, \cite{AGW13}) are the best candidates and will be introduced in our 2D model in future works.

\section*{Acknowledgments}
{D.H. and S.M. thank funding by the European Research Council through ERC SPIRE grant 647383, the CNES PLATO grant at CEA Saclay and the MESA website.}

\bibliographystyle{cs19proc}
\bibliography{mabib.bib}

\end{document}